\documentclass[twocolumn,showpacs,preprintnumbers,aps,amsmath,amssymb]{revtex4}

\usepackage[dvips]{graphicx}
\newtheorem{theorem}{Theorem}
\newtheorem{condition}{Condition}
\newtheorem{corollary}{Corollary}

\newtheorem{lemma}{Lemma}
\newtheorem{proposition}{Proposition}

\newcommand{\ket}[1]{\mbox{$| #1 \rangle$}}
\newcommand{\bra}[1]{\mbox{$\langle #1 |$}}

\newcommand{\ketbra}[2]{\mbox{$| #1 \rangle \langle #2 |$}}
\newcommand{\dket}[1]{\mbox{$| #1 \rangle\rangle$}}
\newcommand{\dbra}[1]{\mbox{$\langle\langle #1 |$}}
\newcommand{\dbraket}[2]{\mbox{$\langle\langle #1 | #2\rangle\rangle$}}
\newcommand{\dketbra}[2]{\mbox{$| #1 \rangle\rangle \langle\langle #2 |$}}

\begin{document}
\preprint{}
\title{On the Matrix Representation of Quantum Operations}
\author{Yoshihiro Nambu and Kazuo Nakamura}
\affiliation{Fundamental and Environmental Research Laboratories, NEC, 34
Miyukigaoka, Tsukuba, Ibaraki 305-8501, Japan}

\begin{abstract}
This paper considers two frequently used matrix representations --- what we call 
the $\chi$- and $\mathcal{S}$-matrices --- of a quantum operation and their 
applications. The matrices defined with respect to an arbitrary operator
basis, that is, the orthonormal basis for the space of linear operators on the
state space are considered for a general operation acting on a single 
or two \textit{d}-level quantum system (qudit). We show that the two matrices are 
given by the expansion coefficients
of the Liouville superoperator as well as the associated bijective, positive
operator on the doubled-space defined with respect to two types of induced
operator basis having different tensor product structures, i.e., Kronecker
products of the relevant operator basis and dyadic products of the associated bipartite state
basis. The explicit conversion formulas between the two matrices are established as
a computable matrix multiplication. Extention to more qudits case is trivial. 
Several applications of these matrices and the conversion formulas in quantum 
information science and technology are presented.

\end{abstract}

\pacs{03.67.-a, 03.65.Wj}
\keywords{quantum, operation, matrix, representation}

\date{\today}

\maketitle

\section{Introduction}
\label{intro}
The formalism of quantum operations offers us a powerful tool for describing the
dynamics of quantum systems occurring in quantum computation, including
unitary evolution as well as non-unitary evolution \cite{Kraus,Nielsen}. It can deal 
with several central issues: measurements in the middle of the computation, 
decoherence and noise, using probabilistic subroutines, etc. \cite{Aharanov}. 
It describes the most general transformation allowed by quantum mechanics for an
initially isolated quantum system \cite{Stelmachovic,Hayashi}. Experimental
characterization and analysis of quantum operations is an essential component
of on-going efforts to develop devices capable of reliable quantum computing
and quantum communications, and is a research subject of considerable recent interest. 
There have been extensive efforts on the statistical
estimation of quantum operations occurring in natural or engineered quantum 
processes from experimental data known as \textquotedblleft quantum channel
identification\textquotedblright\ or \textquotedblleft quantum process
tomography\textquotedblright 
\cite{Nambu,Altepeter,Mitchell,Martini,O'Brien,Secondi,Nambu2}.

There are several ways of introducing the notion of a quantum operation, one of
which is to consider it as superoperators acting on the space of linear
operators \cite{Caves,Aharanov,Tarasov}. Any physical quantum
operation has to be described by a superoperator that is completely positive
(CP); i.e., it should map the set of density operators acting on the trivially
extended Hilbert space to itself \cite{Kraus,Nielsen}. It is known that any CP map can be
decomposed into the so-called Kraus form by a set of Kraus
operators \cite{Kraus}. However, this description is unique only up to unitary
equivalence \cite{Nielsen}, just like the decomposition of a given density operator
into convex sum of distinct, but not necessarily orthogonal, projectors is
unique only up to unitary equivalence \cite{Hughston,Nielsen}. Alternatively,
the superoperators can be represented in matrix form by providing an operator
basis \cite{D'Ariano}, i.e., the orthonormal basis for the space of linear
operators on the state space, just as the operators on the Hilbert space can
be represented in matrix form by providing a state basis for the Hilbert
space. For example, the density operator can be represented by a density matrix
defined with respect to the chosen state basis. 
The density matrix provides a unique description of the quantum state once the
state basis is fixed, although we still have a freedom in choosing the state
basis. Similarly, a quantum operation can also be uniquely described using the
matrix once the operator basis has been fixed.

One can construct many different types of matrix representation. This 
paper considers two different matrix representations for superoperators 
frequently found in the literature. The first one is what we call the $\chi$-matrix,
which is also called the process or dynamical matrix by several
authors \cite{Nielsen,Chuang,Altepeter,O'Brien,Nambu2,Sudarshan,Zyczkowski}. Let
us consider a \textit{d}-dimensional Hilbert space (\textit{H}-space)
$\mathcal{H}_{d}$, and the space of linear operators acting on $\mathcal{H}%
_{d}$ with a scalar product $\langle {\hat{A},\hat{B}}\rangle
\equiv\mathrm{Tr}\hat{A}^{\dag}\hat{B}$, that is, the Hilbert-Schmidt space
(\textit{HS}-space) $\mathcal{HS}_{d}$. If we choose the fixed basis set $\{
{\hat{E}_{\alpha}}\}  _{\alpha=0}^{d^{2}-1}$ in $\mathcal{HS}_{d}$, the
linear operation $\mathfrak{S}$ can be represented by the binary form of the superoperator
\begin{equation}
\mathcal{\hat{\hat{S}}}(\odot)=\sum\limits_{\alpha,\beta
=0}^{d^{2}-1}{\chi_{\alpha\beta}\hat{E}_{\alpha}\odot\hat{E}_{\beta}^{\dag}}
\label{eq1}
\end{equation}
acting on $\mathcal{HS}_{d}$, which maps a linear operator in $\mathcal{HS}
_{d}$ into another one. In Eq. (\ref{eq1}), the substitution symbol $\odot$ should be
replaced by a transformed operator, and double-hat $\Hat{\Hat{ }}$ is used to
distinguish the superoperator from an ordinary operator acting on $\mathcal{H}
_{d}$. The coefficients $\chi_{\alpha\beta}$ form a
$d^{2}\times d^{2}$ positive matrix $\chi\equiv\left[  {\chi_{\alpha\beta}
}\right]  _{\alpha,\beta=0}^{d^{2}-1}$, if $\mathfrak{S}$ is a physical quantum
operation \cite{Arrighi}.

Alternatively, another matrix representation of a quantum operation is given in
terms of the Liouville formalism \cite{Caves,Blum,Royer}. In this formalism,
the linear operators in $\mathcal{HS}_{d}$ are identified with the supervectors
in a Liouville space (\textit{L}-space) $\mathcal{L}_{d^{2}}$. Introducing a
double bra-ket notation for the elements of $\mathcal{L}_{d^{2}}$ \cite{Royer},
we associate every operator $\hat{A}$ with an \textit{L}-ket $\dket{\hat{A}}$ 
and its Hermitian conjugate
operator $\hat{A}^{\dagger}$ with an \textit{L}-bra $\dbra{
\hat{A}}$. The space $\mathcal{L}_{d^{2}}$ is furnished
with an inner product $\dbraket{\hat{A}}{\hat{B}}=\mathrm{{Tr}}\hat{A}^{\dag}\hat{B}$, and
constitutes a $d^{2}$-dimensional Hilbert space. Then by chosing an arbitrary fixed
set of operator basis $\{  {\hat{E}_{\alpha}}\}  _{\alpha=0}%
^{d^{2}-1}$ in $\mathcal{HS}_{d}$, any linear operation $\mathfrak{S}$ can be
written as the superoperator
\begin{equation}
\mathcal{\hat{\hat{S}}}=\sum\limits_{\alpha,\beta=0}^{d^{2}-1}\mathcal{S}%
_{\alpha\beta} \dketbra{\hat{E}_{\alpha}}{\hat{E}_{\beta}} \label{eq2}%
\end{equation}
acting on $\mathcal{L}_{d^{2}}$, which maps a \textit{L}-space supervector
into another one. The coefficients $\mathcal{S}_{\alpha\beta}$ form a
$d^{2}\times d^{2}$ complex matrix $\mathcal{S}\equiv\left[  \mathcal{S}%
{_{\alpha\beta}}\right]  _{\alpha,\beta=0}^{d^{2}-1}$. They represent the
amplitudes of the operator components $\hat{E}_{\alpha}$ contained in the state 
after applying the quantum operation on the operator component $\hat{E}_{\beta}$.
We call it the $\mathcal{S}$-matrix by analogy with the S-matrix appearing in
time-independent scattering theory \cite{Hawking,Wald}. The $\mathcal{S}%
$-matrix has been actually used to describe quantum operations by several
researchers on quantum information science \cite{Mitchell,Martini,Fujiwara}. The $\chi
$- and $\mathcal{S}$-matrices offer us the most general description of the
dynamics of initially isolated quantum systems allowed in quantum mechanics,
just as the density matrix offers us the most general description of the quantum
mechanical state.

The choice of matrix representation type is a matter of convenience,
depending on the application. We will discuss later how the $\chi$- and
$\mathcal{S}$-matrices are useful for the analysis and design of quantum
operations. Although these matrices indeed have their own useful applications,
their mutual relation is non-trivial and has not been clarified. The main
purpose of this paper is to clarify the underlying relation between two
different matrix representations of quantum operations, and to provide the way
for building bridges across the different classes of applications. We here
consider a quantum operation acting on the state of a single \textit{d}-level
quantum system (abbreviated as single-qudit operation) or a two \textit{d}%
-level quantum systems (two-qudit operation). We start in Sec. \ref{operator basis} by recalling 
the notion of operator basis and its properties, which is helpful for the subsequent discussions.
We note the equivalence between the supervectors in the \textit{L}-space
$\mathcal{L}_{d^{2}}$ and the vectors in the doubled Hilbert space
$\mathcal{H}_{d}^{\otimes2}$=$\mathcal{H}_{d}\otimes\mathcal{H}_{d}$. This equivalence 
implies that for any operator-basis set $\{{\hat{E}_{\alpha}}\}_{\alpha=0}^{d^{2}-1}$ 
for $\mathcal{HS}_{d}$, there is an isomorphic state-basis set $\{\dket{\hat{E}_{\alpha}}\}
_{\alpha=0}^{d^{2}-1}$ for$\mathcal{H}_{d}^{\otimes2}$. 
We review several properties of the operator basis for later
discussion. In Sec. \ref{matrix}, we first consider the single-qudit
operations. We show that $\chi$- and $\mathcal{S}$-matrices are given by the
expansion coefficients of the \textit{L}-space superoperator $\mathcal{\hat
{\hat{S}}}$ and the associated operator $\hat{\hat{\chi}}\equiv\mathcal{\hat
{\hat{S}}}\otimes\mathcal{\hat{\hat{I}}}(d\hat{\rho}_{I})$ acting on
$\mathcal{H}_{d}^{\otimes2}$ defined with respect to two types of induced
operator basis on $\mathcal{H}_{d}^{\otimes2}$ having different tensor product
structures. Here, $\hat{\rho}_{I}$ is the density operator of the isotropic
state on $\mathcal{H}_{d}^{\otimes2}$, and $\mathcal{\hat{\hat{I}}}\left(
\odot\right)  =\odot$ is the identity superoperator acting on the
\textit{HS}-space of the second system. This result implies that there is a
bijection between $\mathcal{\hat{\hat{S}}}$ and $\hat{\hat{\chi
}}$, from which we can deduce the conversion formula between $\chi$- and
$\mathcal{S}$-matrices as a computable matrix algebra. Although Nielsen and
Chuang have considered such a formula \cite{Nielsen,Chuang}, their method requires
finding matrix inverses to convert from the $\mathcal{S}$-matrix to the $\chi$-matrix.
Here, we show a conversion formula without matrix inversion. We then 
extend the formula to two-qudit operations. We also briefly
review the requirement for the $\chi$-matrix to represent physical 
operations. In Sec. \ref{applications}, we illustrate the applications
of the present formulation. First, we discuss how $\chi$- and $\mathcal{S}%
$-matrices can be obtained experimentally. We describe a typical procedure to
obtain the $\chi$- and $\mathcal{S}$-matrices defined with respect to an 
arbitrary operator basis set. Next, we discuss how these matrices and the
present conversion formulas are useful for the analysis and design of the
quantum operations, quantum circuits, as well as quantum algorithms. In Sec.
\ref{conclusions}, we summarize the results.

\section{Operator basis}
\label{operator basis}
It has been noted by many researchers that the supervectors defined in the 
\textit{L}-space $\mathcal{L}_{d^{2}}$ can be identified with the vectors in 
the doubled Hilbert space $\mathcal{H}_{d}^{\otimes2}$ 
\cite{Ben-Reuven1,Ben-Reuven2,Ben-Reuven3,Ben-Reuven4}. Let us start by
reviewing this fact, briefly. Consider an arbitrary chosen set of an
orthonormal basis $\{  {\left\vert i\right\rangle }\}  _{i=1}%
^{d-1}$ for $\mathcal{H}_{d}$ (denoted as standard state basis). Any linear
operator $\hat{A}$ in $\mathcal{HS}_{d}$ can be expanded as $\hat{A}%
=\sum\nolimits_{i,j=0}^{d-1}A_{ij}\ketbra{i}{j}$ where the dyadic products 
$\ketbra{i}{j}$ form a basis for $\mathcal{HS}_{d}$. The
\textit{L}-space supervectors corresponding to the dyadic operator 
$\ketbra{i}{j}$ is denoted by the double ket
$\dket{ij}$ , with which the
\textit{L}-space supervector associated with $\hat{A}$ is written as
$\dket{\hat{A}} =\sum
\nolimits_{i,j=0}^{d-1}{A_{ij}\dket{ij}}$. The scalar product of two \textit{L}-space supervectors
$\dket{\hat{A}}$ and $\dket{\hat{B}}$ is defined as
\begin{equation}
\dbraket{\hat{A}}{\hat{B}}
=\mathrm{{Tr}}\hat{A}^{\dag}\hat{B}, \label{eq3}%
\end{equation}
which introduces a metric of an \textit{L}-space. The vectors $\dket{ij}$ form a basis for a 
$d^{2}$-dimensional Hilbert space. Thus, we can safely identify $\dket{ij}$ with the product state $\ket{i} \otimes\ket{j}$. Then, the \textit{L}-space
vector associated with $\hat{A}$ can be identified with vector $\dket{\hat{A}}
\equiv (  {\hat{A}\otimes\hat{I}})  \dket{\hat{I}} $ in the doubled space 
$\mathcal{H}_{d}^{\otimes2} 
$\cite{D'Ariano2,Arrighi}, where $d^{-1/2}\dket{\hat{I}} \equiv d^{-1/2}\sum\nolimits_{i=0}%
^{d-1}\ket{i} \otimes\ket{i}$ is the
isotropic state in $\mathcal{H}_{d}^{\otimes2}$ \cite{Horodecki,Terhal}, 
and $\hat{I}$ is the identity operator in
$\mathcal{HS}_{d}$. It may be helpful to recall the mathematical
representations of $\hat{A}$ in the space $\mathbb{C}^{d\times
d}$ of $d\times d$ complex matrices and the vector $\dket{\hat{A}}$ in $\mathbb{C}^{d^{2}}$. Consider a
representation where $\ket{i}$ is a column vector in
$\mathbb{C}^{d}$ with a unit element in the \textit{j}th row and zeros
elsewhere. Then, $\hat{A}$ is identified with the $d\times d$ complex
matrix $A\equiv[A_{ij}]_{i,j=0}^{d-1}$ in $\mathbb{C}%
^{d\times d}$, and $\dket{\hat{A}}$ is obtained by placing the entries of a $d\times d$ matrix
into a column vector of size $d^{2}$ row-by-row, i.e.,
\begin{equation}
\dket{\hat{A}} \equiv\left[
{A_{11},\cdots,A_{1d},A_{21},\cdots,A_{2d},\cdots,A_{d1},\cdots,A_{dd}%
}\right]  ^{T}. \label{eq4}%
\end{equation}
Therefore, $\dket{\hat{A}}$ contains the same elements as $\hat{A}$ but in 
different positions. This and Eq. (\ref{eq3}) indicate that $\hat{A}$ and 
$\dket{\hat{A}}$ are isometrically isomorphic. Accordingly, we
may identify the \textit{L}-space with the doubled Hilbert space, i.e.,
$\mathcal{L}_{d^{2}}{=H}_{d}^{\otimes2}$. Hereafter, we use a common symbol
$\mathcal{H}_{d}^{\otimes2}$ to denote both these spaces without loss of clarity.

It will be useful for the later discussion to note the following relations
hold:
\begin{equation}
\hat{A}\otimes\hat{B}\dket{\hat{C}}=\dket{\hat{A}\hat{C}\hat{B}^{T}},
\label{eq5}%
\end{equation}%
\begin{equation}
Tr_{2}[\dket{\hat{A}}_{12}{}_{12}\dbra{\hat{B}}]=({\hat{A}\hat{B}^{\dag}})^{(1)}, 
\label{eq6}%
\end{equation}%
\begin{equation}
Tr_{1}[\dket{\hat{A}}_{12}{}_{12}\dbra{\hat{B}}]=({\hat{A}^{T}\hat{B}^{\ast}})^{(2)}, 
\label{eq7}%
\end{equation}
where the indices refer to the factors in $\mathcal{H}_{d}^{\otimes2}$ in
which the corresponding operators have a nontrivial action, and the
transposition and conjugation are referred to the chosen standard state
basis \cite{D'Ariano}.

Now, let us consider the operator basis, that is, the complete basis for
$\mathcal{HS}_{d}$. Consider an arbitrary set of $d^{2}$-vectors in
$\mathcal{H}_{d}^{\otimes2}$. From the above isomorphism, this set can be
written as $\{  {\dket{\hat{E}_{\alpha}}}\}_{\alpha=0}^{d^{2}-1}$, where $\{\hat
{E}_{\alpha}\}_{\alpha=0}^{d^{2}-1}$ is the associated set in
$\mathcal{HS}_{d}$. The set $\{\dket{\hat{E}_{\alpha}}\}_{\alpha=0}^{d^{2}-1}$ is the state
basis for $\mathcal{H}_{d}^{\otimes2}$ iff it is orthonormal, i.e.,
\begin{equation}
\dbraket{\hat{E}_{\alpha}}{\hat{E}_{\beta}}=\delta_{\alpha\beta}, 
\label{eq8}%
\end{equation}
and it is complete, i.e.,
\begin{equation}
\sum\limits_{\alpha=0}^{d^{2}-1}\dketbra{\hat{E}_{\alpha}}{\hat{E}_{\alpha}}=\hat{I}\otimes\hat{I}. 
\label{eq9}%
\end{equation}
The previous discussion shows that the vector $\dket{\hat{A}}$ in $\mathcal{H}_{d}^{\otimes2}$ is identified
with \textit{L}-space supervectors associated with the operator $\hat{A}$ in
$\mathcal{HS}_{d.}$ This implies the following proposition.

\begin{proposition}
\label{P1}
A set of the operators $\{  {\hat{E}_{\alpha}}\}_{\alpha=0}^{d^{2}-1}$ 
is a basis set for $\mathcal{HS}_{d}$ iff a set of states
$\{\dket{\hat{E}_{\alpha}}\}_{\alpha=0}^{d^{2}-1}$ is the basis set for 
$\mathcal{H}_{d}^{\otimes 2}$.
\end{proposition}
We argue below that this is true. First, we note the following lemmmas.

\begin{lemma}
\label{L1}A set of the operators $\{  {\hat{E}_{\alpha}}\}
_{\alpha=0}^{d^{2}-1}$ in $\mathcal{HS}_{d}$ is orthonormal iff a set of
states $\{\dket{\hat{E}_{\alpha}}\}_{\alpha=0}^{d^{2}-1}$ in $\mathcal{H}%
_{d}^{\otimes2}$ is orthonormal.
\end{lemma}
This is a trivial consequence of Eq. (\ref{eq3}).

\begin{lemma}
\label{L2}A set of the operators $\{  {\hat{E}_{\alpha}}\}_{\alpha=0}^{d^{2}-1}$ 
in $\mathcal{HS}_{d}$ is complete iff a set of states $\{\dket{\hat{E}_{\alpha}}\}
_{\alpha=0}^{d^{2}-1}$ in $\mathcal{H}_{d}^{\otimes2}$ is complete.
\end{lemma}
To prove this, the following Theorem is helpful \cite{D'Ariano}:

\begin{theorem}
\label{Th1}
(D'Ariano, Presti, and Sacchi (2000)) A set of the operators
$\{\hat{E}_{\alpha}\}_{\alpha=0}^{d^{2}-1}$\textit{\ in}
$\mathcal{HS}_{d}$ is complete iff it satisfies one of the following
equivalent statements:
\begin{enumerate}
\item For any linear operator $\hat{A}$\textit{ on }$\mathcal{H}_{d}$, we
have
\begin{equation}
\hat{A}=\sum\limits_{\alpha=0}^{d^{2}-1}{(  {\mathrm{{Tr}}\hat{E}%
_{\alpha}^{\dag}\hat{A}})  \hat{E}_{\alpha}.} \label{eq10}%
\end{equation}
\item Let $\hat{\hat{\mathcal{E}}}_{depol}(\cdots)$ be the superoperator on
the space of linear operators in $\mathcal{HS}_{d}$ describing completely
depolarizing operation. For any linear operator $\hat{A}$ on $\mathcal{H}_{d}%
$, we have
\begin{equation}
\hat{\hat{\mathcal{E}}}_{depol}(\hat{A})=\frac{1}{d}(\mathrm{{Tr}}\hat{A})\hat{I}%
=\frac{1}{d}\sum\limits_{\alpha=0}^{d^{2}-1}{\hat{E}_{\alpha}\hat{A}\hat
{E}_{\alpha}^{\dag}.} \label{eq11}%
\end{equation}
\item For chosen any state basis $\{\ket{i}\}_{i=1}^{d-1}$ for $\mathcal{H}_{d}$, we have
\begin{equation}
\sum\limits_{\alpha=0}^{d^{2}-1}{\bra{n} \hat{E}_{\alpha
}^{\dag}\ket{m} \bra{l} \hat{E}_{\alpha}%
}\ket{k}=\delta_{nk}\delta_{ml}. 
\label{eq12}%
\end{equation}
\end{enumerate}
\end{theorem}
Now, let us prove Lemma \ref{L2}. If $\{\dket{\hat
{E}_{\alpha}}\}  _{\alpha=0}^{d^{2}-1}$ is
complete, Eq. (\ref{eq9}) must be satisfied. Then, for any $\hat{A}$
in $\mathcal{HS}_{d}$, we have
\begin{eqnarray*}
(\hat{A}\otimes\hat{I})\dket{\hat{I}}=\dket{\hat{A}}
&=&\sum\limits_{\alpha=0}^{d^{2}-1}
\dket{\hat{E}_{\alpha}} 
\dbraket{\hat{E}_{\alpha}}{\hat{A}}
\\
&=&\sum\limits_{\alpha=0}^{d^{2}-1}(\mathrm{Tr}\hat{E}_{\alpha}^{\dag}\hat{A})  \hat
{E}_{\alpha}\otimes\hat{I}\dket{\hat{I}}.
\end{eqnarray*}
Since this holds for any $\hat{A}$, Eq. (\ref{eq10}) must be satisfied. Hence,
$\{{\hat{E}_{\alpha}}\}_{\alpha=0}^{d^{2}-1}$ is
complete. Conversely, if $\{{\hat{E}_{\alpha}}\}_{\alpha
=0}^{d^{2}-1}$ is complete, then for any $\dket{\hat{A}}$ 
in $\mathcal{H}_{d}^{\otimes2}$, we have
\begin{eqnarray*}
\dket{\hat{A}}=(\hat{A} \otimes \hat{I})\dket{\hat{I}}
&=&\sum\limits_{\alpha=0}^{d^{2}-1}({\mathrm{Tr}\hat{E}_{\alpha}^{\dag}\hat{A}})
\hat{E}_{\alpha}\otimes\hat{I}\dket{\hat{I}}
\\
&=&\sum\limits_{\alpha=0}^{d^{2}-1}
\dket{\hat{E}_{\alpha}}
\dbraket{\hat{E}_{\alpha}}{\hat{A}}.%
\end{eqnarray*}
Since this holds for any $\dket{\hat{A}}$ , Eq. (\ref{eq9}) must be satisfied. 
Hence, $\{\dket{\hat{E}_{\alpha}} \}_{\alpha=0}^{d^{2}-1}$ is complete.

From these two lemmas, we obtain Proposition \ref{P1}. This indicates that the state 
basis $\dket{\hat E_{\alpha}}$ for $\mathcal{H}_{d}^{\otimes2}$ has bijective correspondence
to the operator basis $\hat E_{\alpha}$ for $\mathcal{HS}_{d}$. The following
corollary is a consequence of Theorem \ref{Th1}, which will be useful for the
later discussions.

\begin{corollary}
\label{C1}Let $\{\hat{E}_{\alpha}\}_{\alpha=0}^{d^{2}-1}$ 
be an arbitrary chosen set of an operator basis for $\mathcal{HS}%
_{d}$. The isotropic state in $\mathcal{H}_{d}^{\otimes2}$ is written as
\begin{equation}
\hat{\rho}_{I}=\frac{1}{d}\dketbra{\hat{I}}{\hat{I}}
=\frac{1}{d}\sum\limits_{\alpha=0}^{d^{2}-1}\hat{E}_{\alpha}\otimes\hat
{E}_{\alpha}^{\ast}, \label{eq13}%
\end{equation}
and the swap operator $\hat{V}$\textit{on} $\mathcal{H}_{d}^{\otimes2}$ is
written as
\begin{equation}
\hat{V}=\sum\limits_{\alpha=0}^{d^{2}-1}\hat{E}_{\alpha}\otimes\hat{E}%
_{\alpha}^{\dag}. \label{eq14}%
\end{equation}
\end{corollary}
Equation (\ref{eq13}) can be proven by explicit evaluation of the matrix
elements and using Eq. (\ref{eq12}). Equation (\ref{eq14}) is obtained by
performing the partial transpose on both sides of Eq. (\ref{eq13}) with respect to
the second system.\\

\textbf{Examples of the operator basis}

We show three illustrative examples of the operator basis for $\mathcal{H}%
_{d}$ frequently found in the literature. The first example is a set of
transition operators $\hat{\pi}_{\left(  {i,j}\right)  }:=\ketbra{i}{j}$ 
with $i,j=0,\cdots,d-1$, and
$(i,j):=di+j$ \cite{Mahler}. The associated states form a basis set 
$\{\dket{\hat{\pi}_{\alpha}}\}_{\alpha=0}^{d^{2}-1}$ whose elements are the 
tensor product of the standard state basis, i.e., $\dket{\hat{\pi}_{({i,j})}}
=\ket{i}\otimes\ket{j}$. The next example is a set of unitary
irreducible representations of the group \textit{SU}(\textit{d}) or the
discrete displacement operators on the phase-space torus, whose elements are
$\hat{U}_{(m,n)}={{\omega^{mn/2}\sum\nolimits_{k=0}%
^{d-1}{\omega^{mk}\hat{\pi}_{(k\oplus n,k)  }/}}}\sqrt{d}$ with
$m,n=0,\cdots,d-1$, where $\omega=1^{1/d}=e^{i2\pi/d}$ and $\oplus$ 
denotes addition modulo $d$ \cite{Mahler,Miquel,Aolita}. The associated states
form a basis set $\{\dket{\hat{U}_{\alpha}}\}_{\alpha=0}^{d^{2}-1}$ of $d^{2}%
$-orthogonal maximally entangled states in $\mathcal{H}_{d}^{\otimes2}$. The
last example is a set of $d^{2}-1$ traceless Hermitian generators of the group
\textit{SU}(\textit{d}) supplemented with the normalized identity operator given
by $\{\hat{\lambda}_{\alpha}\}_{\alpha=0}^{d^2-1}
=\{\hat{I}/\sqrt{d},\hat{u}_{0,1},\hat{u}_{0,2},\cdots,\hat{u}_{d-2,d-1},
\hat{v}_{0,1},\hat{v}_{0,2},\cdots,\hat{v}_{d-2,d-1},\\
\hat{w}_{1},\hat{w}_{2},\cdots,\hat{w}_{d-1}\}$ where $d(d-1)$ off-diagonal
generators are given by $\hat{u}_{i,j}={{(  {\hat{\pi}_{(
i,j)}+\hat{\pi}_{(j,i)}})}}/\sqrt{2}$,
$\hat{v}_{i,j}={{i({\hat{\pi}_{(i,j)}-\hat{\pi
}_{(j,i)}})}/\sqrt{2}}$ with $0\leq i<j\leq d-1$, and $d-1$ diagonal 
generators are given by $\hat{w}_{k}={{(  {-\sum
\nolimits_{i=0}^{k-1}{\hat{\pi}_{(  {i,i})  }}+k\hat{\pi}_{(
{k,k})  }})  /}}\sqrt{k(k+1)}$ with $1\leq k\leq d-1$ \cite{Mahler}. 
The choice of basis is of course a matter of convenience, depending on the application.

Since the associated sets $\{\dket{\hat{\pi}_{\alpha}}\}_{\alpha=0}^{d^{2}-1}$, 
$\{\dket{\hat{U}_{\alpha}}\}_{\alpha=0}^{d^{2}-1}$, and $\{\dket{\hat{\lambda
}_{\alpha}}\}_{\alpha=0}^{d^{2}-1}$ are
state bases for $\mathcal{H}_{d}^{\otimes2}$, they should be unitarily related. 
This implies that the operator bases ${\hat{\pi
}_{\alpha}}$, ${\hat{U}_{\alpha}}$, and ${\hat{\lambda}_{\alpha}}$ should also be
unitarily related. In general, two sets of state basis $\{\dket{\hat{E}_{\alpha}}\}_{\alpha
=0}^{d^{2}-1}$ and $\{\dket{\hat{F}_{\alpha}}\}_{\alpha=0}^{d^{2}-1}$ in
$\mathcal{H}_{d}^{\otimes2}$ are unitarily related, i.e.,
\begin{equation}
\dket{\hat{F}_{\beta}}
=\sum\limits_{\alpha=0}^{d^{2}-1}
\dket{\hat{E}_{\alpha}}
\dbraket{\hat{E}_{\alpha}}{\hat{F}_{\beta}}
=\sum\limits_{\alpha=0}^{d^{2}-1}\dket{\hat{E}_{\alpha}}\mathcal{U}{_{\alpha\beta}} 
\label{eq15}
\end{equation}
iff the operator bases ${\hat{E}_{\alpha}}$ and ${\hat{F}_{\alpha}}$ in
$\mathcal{HS}_{d}$ are unitarily related, i.e.,
\begin{equation}
\hat{F}_{\beta}=\sum\limits_{\alpha=0}^{d^{2}-1}{\hat{E}_{\alpha}}%
\mathcal{U}{_{\alpha\beta}.} \label{eq16}%
\end{equation}
In Eqs. (\ref{eq15}) and (\ref{eq16}), $\mathcal{U}_{\alpha\beta}=\dbraket
{\hat{E}_{\alpha}}{\hat{F}_{\beta}}
=\mathrm{Tr}\hat{E}_{\alpha}^{\dag}\hat{F}_{\beta}$ is a $\alpha\beta$-entry
of the $d^{2}\times d^{2}$ unitary matrix $\mathcal{U}$. If we consider a 
unitary superoperator acting on the vectors in $\mathcal{H}_{d}^{\otimes2}$,
\begin{equation}
\mathcal{\hat{\hat{U}}}=\sum\limits_{\alpha,\beta=0}^{d^{2}-1}\mathcal{U}%
{_{\alpha\beta}
\dketbra{\hat{E}_{\alpha}}{\hat{E}_{\beta}},} 
\label{eq17}%
\end{equation}
Equations. (\ref{eq15}) and (\ref{eq16}) are the unitary
transformation of the operators. Note that Eq. (\ref{eq16})
does not imply unitary equivalence of $\hat{E}_{\alpha}$ and ${\hat{F}_{\alpha}%
}$, i.e., $\hat{F}_{\beta}=\hat{W}\hat{E}_{\alpha}\hat{W}^{\dag}$ for some
unitary operator $\hat{W}$ in $\mathcal{HS}_{d}$, although the unitary equivalence
of $\hat{E}_{\alpha}$ and ${\hat{F}_{\alpha}}$ implies Eq. (\ref{eq16}). 
In general, $\hat{F}_{\beta}\neq\hat{W}\hat{E}_{\alpha}\hat
{W}^{\dag}$ for any unitary operator $\hat{W}$ in $\mathcal{HS}_{d}$, even if
Eq. (\ref{eq16}) holds. For example, ${\hat{\pi}_{\left(
{i,j}\right)  }}$ is the operator basis of all the elements of which have rank
one, whereas ${\hat{U}_{\alpha}}$ and ${\hat{\lambda}_{\alpha}}$ are those
operator bases of all the elements of which have rank exceeding one. Therefore,
$\hat{\pi}_{({i,j})}$ is induced by a standard state basis
$\{\ket{i}\}_{i=0}^{d-1}$ for
$\mathcal{H}_{d}$, whereas ${\hat{U}_{\alpha}}$ and ${\hat{\lambda
}_{\alpha}}$ are not. This clearly indicates that either the set $\{
\hat{U}_{\alpha}\}_{\alpha=0}^{d^{2}-1}$ or $\{\hat{\lambda
}_{\alpha}\}_{\alpha=0}^{d^{2}-1}$ is unitarily related to the set 
$\{\hat{\pi}_{\alpha}\}_{\alpha=0}^{d^{2}-1}$, but is not unitarily equivalent to this set.

\section{Matrix representation of quantum operations}
\label{matrix}
Let us turn our attention to a single-qudit operation. As shown
in Sec. \ref{intro}, this quantum operation can be represented by either an 
\textit{HS}-space superoperator or an \textit{L}-space superoperator, 
in which the $\chi$- and $\mathcal{S}$-matrices are introduced
with respect to the arbitrary, but associated sets of basis $\{  {\hat
{E}_{\alpha}}\}  _{\alpha=0}^{d^{2}-1}$ for $\mathcal{HS}_{d}$ and
$\{\dket{\hat{E}_{\alpha}}\}_{\alpha=0}^{d^{2}-1}$ for $\mathcal{H}_{d}^{\otimes2}$,
respectively. In this section, the underlying relationship between these two
matrix representations is discussed.

Let us first consider the \textit{L}-space superoperator.
In the Liouville formalism, the operators in $\mathcal{HS}_{d}$ are identified
with the vectors in $\mathcal{H}_{d}^{\otimes2}$. Any operation
$\mathfrak{S}$ is identified with the one-sided operator $\mathcal{\hat
{\hat{S}}}$ acting on $\mathcal{H}_{d}^{\otimes2}$, which can be expanded
using the state basis $\dket{\hat{E}_{\alpha}}$ for $\mathcal{H}_{d}^{\otimes2}$ 
as shown in Eq. (\ref{eq2}).
The elements of a $d^{2}\times d^{2}$ matrix $\mathcal{S}$ are formally
written as $\mathcal{S}_{\alpha\beta}=\dbra{\hat{E}_{\alpha}} \mathcal{\hat{\hat{S}}}
\dket{\hat{E}_{\beta}}$. Alternatively, the same operation
$\mathfrak{S}$ is written as a two-sided superoperator acting on
the operator in $\mathcal{HS}_{d}$, which can be expanded
using the operator basis $\hat{E}_{\alpha}$ for $\mathcal{HS}_{d}$ 
as shown in Eq. (\ref{eq1}). Since 
$\mathcal{\hat{\hat{S}}} \dket{\hat{E}_{\beta}}=\dket{\mathcal{\hat{\hat{S}}}(\hat{E}_{\beta})}$, 
we find the matrix element $\mathcal{S}_{\alpha\beta}$ can also be written as
\begin{equation}
\mathcal{S}_{\alpha\beta}=
\dbraket{\hat{E}_{\alpha}}{\mathcal{\hat{\hat{S}}}(\hat{E}_{\beta})}
=\sum\limits_{\gamma,\delta=0}^{d^{2}-1}{\chi_{\gamma\delta}
\dbra{\hat{E}_{\alpha}} \hat{E}_{\gamma}\otimes\hat{E}_{\delta}^{\ast}
\dket{\hat{E}_{\beta}},} \label{eq18}%
\end{equation}
where we used Eq. (\ref{eq5}). 
Substituting the right-hand side of Eq. (\ref{eq18}) for $\mathcal{S}%
_{\alpha\beta}$ in Eq. (\ref{eq2}), we find that $\mathcal{\hat{\hat{S}}}$ can
be written in terms of either the matrix $\mathcal{S}$ or $\chi$ as
\begin{equation}
\mathcal{\hat{\hat{S}}}=\sum\limits_{\alpha,\beta=0}^{d^{2}-1}\mathcal{S}%
{_{\alpha\beta}
\dketbra{\hat{E}_{\alpha}}{\hat{E}_{\beta}} }=\sum\limits_{\alpha,\beta=0}^{d^{2}-1}{\chi_{\alpha\beta}%
\hat{E}_{\alpha}\otimes\hat{E}_{\beta}^{\ast}.} \label{eq19}%
\end{equation}
In Eq. (\ref{eq19}), we find two types of induced operator basis on $\mathcal{H}%
_{d}^{\otimes2}$ having different tensor product structures, that is,
Kronecker products $\hat{E}_{\alpha}\otimes\hat{E}_{\beta}^{\ast}$ and dyadic
products $\dketbra{\hat{E}_{\alpha}}{\hat{E}_{\beta}} $ of the state
basis associated with the operator basis set $\{  {\hat{E}_{\alpha}%
}\}_{\alpha=0}^{d^{2}-1}$. Note that both types of basis set do
not cover all the possible basis sets on $\mathcal{H}%
_{d}^{\otimes2}$. Obviously, the former type of basis set covers only those
sets that are factorable with respect to the original and extended system
spaces. For example, the set of $d^{4}$-dyadic products of the $d^{2}%
$-maximally-entangled states in $\mathcal{H}_{d}^{\otimes2}$ is not a
factorable basis set, and can not be covered by the former type.
Similarly, only an operator basis on $\mathcal{H}_{d}^{\otimes2}$ with all 
elements of which have rank one can be reduced to the latter type of the basis
set. Therefore, each type of basis set can describe its own particular subset
of all the possible basis sets on $\mathcal{H}_{d}^{\otimes2}$.

Let us next consider another operator $\hat{\hat{\chi}}$ on $\mathcal{H}%
_{d}^{\otimes2}$, which we call the Choi operator \cite{Choi}. We will show that
this operator has bijective correspondence to the \textit{L}-space
superoperator $\mathcal{\hat{\hat{S}}}$. It is known that the isomorphism
between the operator in $\mathcal{HS}_{d}$ and the bipartite vector in
$\mathcal{H}_{d}^{\otimes2}$ can be straightforwardly extended to the isomorphism
between the superoperator acting on $\mathcal{HS}_{d}$ and the operator acting
on $\mathcal{H}_{d}^{\otimes2}$. Jamio\l kowski first showed that the map
between the \textit{HS}-space superoperator $\mathcal{\hat{\hat{S}}}\left(
\odot\right)$ and the operator $\hat{\hat{\chi
}}\equiv\mathcal{\hat{\hat{S}}}\otimes\mathcal{\hat{\hat{I}}}(d\hat{\rho}%
_{I})$ acting on $\mathcal{H}_{d}^{\otimes2}$ is an isomorphism, where
$\mathcal{\hat{\hat{S}}}\left(  \odot\right)  $ and $\mathcal{\hat{\hat{I}}%
}\left(  \odot\right)  $ act on the \textit{HS}-space of the first and second
systems, respectively \cite{Jamiolkowski}. If we note Eq. (\ref{eq13}) and the
following equivalent relation that follows from Eq. (\ref{eq19})
\[
\mathcal{\hat{\hat{S}}} \dket{\hat{E}_{\beta}} =\sum\limits_{\alpha=0}^{d^{2}-1}
\mathcal{S}{_{\alpha\beta} \dket{\hat{E}_{\alpha}}}\leftrightarrow
\mathcal{\hat{\hat{S}}}(  {\hat{E}_{\beta}})
=\sum\limits_{\alpha=0}^{d^{2}-1}\mathcal{S}{_{\alpha\beta}\hat{E}_{\alpha},}%
\]
it is easy to confirm that $\hat{\hat{\chi}}$ can be written
as
\begin{equation}
\hat{\hat{\chi}}=\sum\limits_{\alpha,\beta=0}^{d^{2}-1}{\chi_{\alpha\beta}
\dketbra{\hat{E}_{\alpha}}{\hat{E}_{\beta}} }%
=\sum\limits_{\alpha,\beta=0}^{d^{2}-1}\mathcal{S}{_{\alpha\beta}\hat
{E}_{\alpha}\otimes\hat{E}_{\beta}^{\ast}.} \label{eq20}%
\end{equation}
Equations (\ref{eq19}) and (\ref{eq20}) are one of the main results of this paper.
From these equations, we find that $\mathcal{\hat{\hat{S}}}$ and $\hat
{\hat{\chi}}$ are complementary to each other in the sense that they can be 
interchanged if we exchange the two operator bases $\dketbra{\hat{E}_{\alpha}}
{\hat{E}_{\beta}} $ and $\hat{E}_{\alpha}\otimes\hat{E}%
_{\beta}^{\ast}$ on $\mathcal{H}_{d}^{\otimes2}$ in their expressions. These
equations show that the $\mathcal{S}$-matrix ($\chi$-matrix) is given by the
expansion coefficients of $\mathcal{\hat{\hat{S}}}$ ($\hat{\hat{\chi}}$) with
respect to $\dketbra{\hat{E}_{\alpha}}{\hat{E}_{\beta}} $ as well as
those of $\hat{\hat{\chi}}$ ($\mathcal{\hat{\hat{S}}}$) with respect to
$\hat{E}_{\alpha}\otimes\hat{E}_{\beta}^{\ast}$, which is explicitly written
as
\begin{equation}
\chi_{\alpha\beta}=\mathrm{Tr}(\dketbra{\hat{E}_{\alpha}}{\hat{E}_{\beta}})
^{\dag}\hat{\hat{\chi}}=\mathrm{Tr}({\hat{E}_{\alpha}\otimes\hat{E}_{\beta}^{\ast}})
^{\dag}\mathcal{\hat{\hat{S}}}, 
\label{eq21}
\end{equation}
\begin{equation}
\mathcal{S}_{\alpha\beta}=\mathrm{Tr}(\dketbra{\hat{E}_{\alpha}}
{\hat{E}_{\beta}})^{\dag}\mathcal{\hat{\hat{S}}}
=\mathrm{Tr}(\hat{E}_{\alpha}\otimes\hat{E}_{\beta}^{\ast})^{\dag}\hat{\hat{\chi}}. 
\label{eq22}
\end{equation}
From Eqs. (\ref{eq19}) and (\ref{eq20}), we can explore the mutual conversion
formulas between $\chi$- and $\mathcal{S}$-matrices. To this end, let us define
a bijection between the two operators on $\mathcal{H}_{d}^{\otimes2}$
originally found by Havel \cite{Havel}.
\begin{equation}
\Lambda(\odot)=\sum\limits_{\gamma=0}^{d^{2}-1}{(\hat{I}\otimes
\hat{\pi}_{\gamma})  \odot (\hat{\pi}_{\gamma}\otimes\hat{I})  ,} 
\label{eq23}%
\end{equation}
which is also considered to be the super-superoperator acting on $\mathcal{H}%
_{d}.$ Then, we have the following Theorem \cite{Havel}.

\begin{theorem}
\label{Th2} (Havel (2003)) For arbitrary operators $\hat{X}$ and $\hat{Y}$ in
$\mathcal{HS}_{d}$, we have
\begin{equation}
\dketbra{\hat{X}}{\hat{Y}}=\Lambda(\hat{X}\otimes\hat{Y}^{\ast}), 
\label{eq24}%
\end{equation}%
\begin{equation}
\hat{X}\otimes\hat{Y}^{\ast}=\Lambda(\dketbra{\hat{X}}{\hat{Y}}). 
\label{eq25}%
\end{equation}
\end{theorem}
Theorem \ref{Th2} connects two relevant operators on $\mathcal{H}_{d}%
^{\otimes2}$ having different tensor product structures, i.e., the Kronecker
product ${\hat{X}\otimes\hat{Y}^{\ast}}$ and the dyadic product $\dketbra{\hat{X}}
{\hat{Y}} $. To prove the Theorem \ref{Th2}, we first note the following lemma.

\begin{lemma}
\label{L3}
The identity operator on $\mathcal{H}_{d}^{\otimes2}$ and the
(unnormalized) density operator of the isotropic state on $\mathcal{H}%
_{d}^{\otimes2}$ are related as follows.
\begin{equation}
\dketbra{\hat{I}}{\hat{I}}=\Lambda(\hat{I}\otimes\hat{I}), 
\label{eq26}%
\end{equation}%
\begin{equation}
\hat{I}\otimes\hat{I}=\Lambda(\dketbra{\hat{I}}{\hat{I}}). 
\label{eq27}%
\end{equation}
\end{lemma}
It is straightforward to confirm Eqs. (\ref{eq26}) and (\ref{eq27}) by writing
$\dketbra{\hat{I}}{\hat{I}} $ and $\hat{I}\otimes\hat{I}$
using the standard state basis $\ket{i}$ explicitly.
Then, it follows that\textit{ }%
\begin{eqnarray*}
\dketbra{\hat{X}}{\hat{Y}}
&=&(\hat{X}\otimes\hat{I}) \dketbra{\hat{I}}{\hat{I}}(\hat{I}\otimes\hat{Y}^{\ast})
\\
&=&\Lambda((\hat{X}\otimes\hat{I})(\hat{I}\otimes\hat{I})(\hat{I}\otimes\hat{Y}^{\ast}))
=\Lambda(\hat{X}\otimes\hat{Y}^{\ast}),
\end{eqnarray*}
\begin{eqnarray*}
\hat{X}\otimes\hat{Y}^{\ast}
&=&(\hat{X}\otimes\hat{I})(\hat{I}\otimes\hat{I})(\hat{I}\otimes\hat{Y}^{\ast})
\\
&=&\Lambda((\hat{X}\otimes\hat{I})  \dketbra{\hat{I}}{\hat{I}}(\hat{I}\otimes\hat{Y}^{\ast}))
=\Lambda(\ketbra{\hat{X}}{\hat{Y}}),
\end{eqnarray*}
where we used Eq. (\ref{eq5}). Accordingly, Theorem \ref{Th2} is proved. At
this point, we note that the action of the bijection $\Lambda\left(  {\odot
}\right)$ corresponds to reshuffling of the matrix introduced by
\.{Z}yczkowski and Bengtsson: if we consider the matrix for the operator on
$\mathcal{H}_{d}^{\otimes2}$ defined with respect to the standard state basis, 
the mapped operator by $\Lambda\left(  {\odot}\right)  $ has a 
reshuffled one of the original matrix \cite{Zyczkowski}. The bijection
$\Lambda(\odot)$ is also closely related to the matrix
realignment introduced by Chen and Wu to discuss the separability criterion for
the bipartite density matrix \cite{Chen}. It is easy to confirm that
$\Lambda(\odot)$ is involutory, that is, $\Lambda
(\Lambda(\odot))=\odot$. It should be also noted that
$\Lambda(\odot)$ does not preserve Hermiticity and the 
rank of the transformed operator, so its spectrum is not
preserved \cite{Zyczkowski}. This means that $\Lambda(\odot)$
represents a non-physical operation. It follows from Theorem \ref{Th2}
that $\hat{\hat{\chi}}=\Lambda(\mathcal{\hat{\hat{S}}})$ and
$\mathcal{\hat{\hat{S}}}=\Lambda(\hat{\hat{\chi}})$, i.e., $\mathcal{\hat{\hat{S}}}$ 
and $\hat{\hat{\chi}}$ are bijective.

From this bijective relation, we can explore the bijection between $\chi$- and $\mathcal{S}%
$-matrices. To this end, we expand $\dketbra{\hat{E}_{\alpha}}{\hat{E}_{\beta}}$ 
in terms of ${\hat{E}_{\alpha}\otimes\hat{E}_{\beta
}^{\ast}}$, and vice versa. Since they are bijective, it follows that
\begin{eqnarray}
\hat{E}_{\alpha}\otimes\hat{E}_{\beta}^{\ast}
&=&\Lambda(\dketbra{\hat{E}_{\alpha}}{\hat{E}_{\beta}})
\nonumber\\
&=&\sum\limits_{\alpha^{\prime},\beta^{\prime},\gamma=0}^{d^{2}-1}
{\dketbra{\hat{E}_{\alpha^{\prime}}}{\hat{E}_{\beta^{\prime}}} M_{\alpha
^{\prime}\beta^{\prime},\alpha\beta},} \label{eq28}%
\end{eqnarray}
\begin{eqnarray}
\dketbra{\hat{E}_{\alpha}}{\hat{E}_{\beta}}
&=&\Lambda(\hat{E}_{\alpha}\otimes\hat{E}_{\beta}^{\ast})
\nonumber\\
&=&\sum\limits_{\alpha^{\prime},\beta^{\prime}=0}^{d^{2}-1}{\hat{E}%
_{\alpha^{\prime}}\otimes\hat{E}_{\beta^{\prime}}^{\ast}
(M^{\dag})_{\alpha^{\prime}\beta^{\prime},\alpha\beta},} 
\label{eq29}%
\end{eqnarray}
where ${M_{\alpha^{\prime}\beta^{\prime},\alpha\beta}}$ is $\alpha^{\prime
}\beta^{\prime};\alpha\beta$-entry of the $d^{4}\times d^{4}$ complex matrix
$M$. It is explicitly given as
\begin{eqnarray}
M_{\alpha^{\prime}\beta^{\prime};\alpha\beta}
&=&\mathrm{Tr}(\dketbra{\hat{E}_{\alpha^\prime}}{\hat{E}_{\beta^\prime}})
^{\dag}\hat{E}_{\alpha}\otimes\hat{E}_{\beta}^{\ast}
\nonumber\\
&=&\sum\limits_{\gamma=0}^{d^{2}-1}{Q_{\alpha^{\prime}\alpha}^{\gamma}R_{\beta\beta^{\prime}}^{\gamma},} 
\label{eq30}%
\end{eqnarray}
where the coefficients%
\begin{equation}
Q_{\alpha\beta}^{\gamma}=\dbra{\hat{E}_{\alpha}}(\hat{I}\otimes\hat{\pi}_{\gamma}) 
\dket{\hat{E}_{\beta}}, \label{eq31}%
\end{equation}%
\begin{equation}
R_{\alpha\beta}^{\gamma}=\dbra{\hat{E}_{\alpha}}(\hat{\pi}_{\gamma}\otimes\hat{I})
\dket{\hat{E}_{\beta}} 
\label{eq32}%
\end{equation}
are the computable matrix elements of the operators ${\hat{I}\otimes\hat{\pi
}_{\gamma}}$ and ${\hat{\pi}_{\gamma}\otimes\hat{I}}$ on $\mathcal{H}%
_{d}^{\otimes2}$ defined with respect to the basis set $\{\dket{\hat{E}_{\alpha}}\}_{\alpha
=0}^{d^{2}-1}$. These coefficients form the 
$d^{2}\times d^{2}$ complex matrices
$Q^{\gamma}\equiv[Q_{\alpha\beta}^{\gamma}]_{\alpha,\beta=0}^{d^{2}-1}$
and $R^{\gamma}\equiv[R_{\alpha\beta}^{\gamma}]_{\alpha,\beta=0}^{d^{2}-1}$. 
It is easy to confirm that $M$ is Hermitian as well as unitary, i.e., 
$(M^{\dag})_{\alpha^{\prime}\beta^{\prime},\alpha\beta}=(M_{\alpha\beta,\alpha^{\prime}\beta^{\prime}})^{\ast}={M_{\alpha^{\prime}\beta^{\prime},\alpha\beta}}$ and 
$\sum\nolimits_{\alpha^{\prime\prime},\beta^{\prime\prime}}
(M^{\dag})_{\alpha\beta,\alpha^{\prime\prime}\beta^{\prime\prime}}
M_{\alpha^{\prime\prime}\beta^{\prime\prime};\alpha^{\prime}\beta^{\prime}}
=\sum\nolimits_{\alpha^{\prime\prime},\beta^{\prime\prime}}
M_{\alpha\beta;\alpha^{\prime\prime}\beta^{\prime\prime}}
(M^{\dag})_{\alpha^{\prime\prime}\beta^{\prime\prime},\alpha^{\prime}\beta^{\prime}}
=\delta_{\alpha\alpha^{\prime}}\delta_{\beta\beta^{\prime}}$. By using these results, we
obtain a bijection between the $\chi$- and $\mathcal{S}$ -matrices:
\begin{equation}
\chi=\sum\limits_{\gamma=0}^{d^{2}-1}{Q^{\gamma}\mathcal{S}R^{\gamma},}
\label{eq33}
\end{equation}
\begin{equation}
\mathcal{S}=\sum\limits_{\gamma=0}^{d^{2}-1}{Q^{\gamma}\chi R^{\gamma},}
\label{eq34}
\end{equation}
which are given by the sum of the multiplication of three known matrices $Q^{\gamma}$, 
$\chi$, and $R^{\gamma}$, and evidently computable.

The above formulation can be straightforwardly extended to describe two-qudit
operations. In this case, different choices of basis sets are allowed 
for two systems. Let us choose the set $\{\hat{E}_{\alpha}\}_{\alpha=0}^{d^{2}-1}$ acting on the
first qudit space and the set $\{\hat{F}_{\alpha}\}_{\alpha=0}^{d^{2}-1}$ 
acting on the second qudit space. The general two-qudit superoperator can be written as
\begin{equation}
\mathcal{\hat{\hat{S}}}
=\sum\limits_{\alpha,\beta,\gamma,\delta=0}^{d^{2}-1}\mathcal{S}_{\alpha\beta,\gamma\delta}
\dket{\hat{E}_{\alpha}}
\dketbra{\hat{F}_{\beta}}{\hat{E}_{\gamma}}
\dbra{\hat{F}_{\delta}} 
\label{eq35}%
\end{equation}
acting on the $d^{4}$-dimensional \textit{L}-space $\mathcal{L}_{d^{4}}$ as
well as
\begin{equation}
\mathcal{\hat{\hat{S}}}\left(  \odot\right)  =\sum\limits_{\alpha,\beta
,\gamma,\delta=0}^{d^{2}-1}{\chi_{\alpha\beta,\gamma\delta}\hat{E}_{\alpha
}\otimes\hat{F}_{\beta}\odot\hat{E}_{\gamma}^{\dag}\otimes\hat{F}_{\delta
}^{\dag}} \label{eq36}
\end{equation}
acting on $\mathcal{HS}_{d^{2}}$. These superoperators are characterized with
the $d^{4}\times d^{4}$ matrices $\mathcal{S}\equiv\left[  \mathcal{S}%
{_{\alpha\beta,\gamma\delta}}\right]  _{\alpha,\beta,\gamma,\delta=0}%
^{d^{2}-1}$ and $\chi\equiv\left[  {\chi_{\alpha\beta,\gamma\delta}}\right]
_{\alpha,\beta,\gamma,\delta=0}^{d^{2}-1}$. The bijective Choi operator is
defined on the $d^{4}$-dimensional \textit{H}-space $\mathcal{H}_{d}%
^{\otimes4}$ that is identified with the \textit{L}-space $\mathcal{L}_{d^{4}%
}$ as follows:
\begin{equation}
\hat{\hat{\chi}}\equiv\mathcal{\hat{\hat{S}}}^{(13)}\otimes\mathcal{\hat
{\hat{I}}}^{(24)}(d^{2}\hat{\rho}_{I}^{(12)}\otimes\hat{\rho}_{I}^{(34)}),
\label{eq37}
\end{equation}
where the indices refer to the factors in $\mathcal{H}_{d}^{\otimes4}$ in
which the corresponding operations have a nontrivial
action \cite{Cirac,Dur,Harrow}. Then, it is straightforward to show that 
$\mathcal{\hat{\hat{S}}}$ and $\hat{\hat{\chi}}$ can be written as follows:
\begin{eqnarray}
\mathcal{\hat{\hat{S}}}
&=&\sum\limits_{\alpha,\beta,\gamma,\delta=0}^{d^{2}-1}\mathcal{S}_{\alpha\beta,\gamma\delta}
\dket{\hat{E}_{\alpha}}
\dketbra{\hat{F}_{\beta}}{\hat{E}_{\gamma}}
\dbra{\hat{F}_{\delta}}
\nonumber\\
&=&\sum\limits_{\alpha,\beta,\gamma,\delta=0}^{d^{2}-1}%
{\chi_{\alpha\beta,\gamma\delta}\hat{E}_{\alpha}\otimes\hat{E}_{\gamma}^{\ast
}\otimes\hat{F}_{\beta}}\otimes\hat{F}_{\delta}^{\ast}. 
\label{eq38}%
\end{eqnarray}
\begin{eqnarray}
\hat{\hat{\chi}}
&=&\sum\limits_{\alpha,\beta,\gamma,\delta=0}^{d^{2}-1}%
{{\chi_{\alpha\beta,\gamma\delta}}}
\dket{\hat{E}_{\alpha}}
\dketbra{\hat{F}_{\beta}}{\hat{E}_{\gamma}}
\dbra{\hat{F}_{\delta}}
\nonumber\\
&=&\sum\limits_{\alpha,\beta,\gamma,\delta=0}^{d^{2}-1}%
\mathcal{S}{_{\alpha\beta,\gamma\delta}\hat{E}_{\alpha}\otimes\hat{E}_{\gamma
}^{\ast}\otimes\hat{F}_{\beta}}\otimes\hat{F}_{\delta}^{\ast}, 
\label{eq39}%
\end{eqnarray}
It follows that the two operators are certainly bijective: $\hat{\hat{\chi}%
}=\Lambda\otimes\Lambda(  \mathcal{\hat{\hat{S}}})  $ and
$\mathcal{\hat{\hat{S}}=}\Lambda\otimes\Lambda(  \hat{\hat{\chi}})$. 
From these bijective relations, we can explore the bijection between the $\chi$- and
$\mathcal{S}$-matrices for a two-qudit operation as
\begin{equation}
\chi=\sum\limits_{\gamma,\lambda=0}^{d^{2}-1}{Q^{\gamma}\otimes S^{\lambda}%
}\mathcal{S}{R^{\gamma}\otimes T^{\lambda},} \label{eq40}%
\end{equation}%
\begin{equation}
\mathcal{S}=\sum\limits_{\gamma,\lambda=0}^{d^{2}-1}{Q^{\gamma}\otimes
S^{\lambda}\chi R^{\gamma}\otimes T^{\lambda},} \label{eq41}%
\end{equation}
where $d^{2}\times d^{2}$ matrices ${Q^{\gamma}}$ and ${R^{\gamma}}$ are given
by Eqs. (\ref{eq31}) and (\ref{eq32}), and the matrix entries of ${S^{\lambda
}}$ and ${T^{\lambda}}$ are given by the matrix elements of the operators
${\hat{I}\otimes\hat{\pi}_{\gamma}}$ and ${\hat{\pi}_{\gamma}\otimes\hat{I}}$
on $\mathcal{H}_{d}^{\otimes2}$ defined with respect to the basis set
$\{\dket{\hat{F}_{\alpha}}\}_{\alpha=0}^{d^{2}-1}$:
\begin{equation}
S_{\alpha\beta}^{\gamma}=\dbra{\hat{F}_{\alpha}}
(\hat{I}\otimes\hat{\pi}_{\gamma})
\dket{\hat{F}_{\beta}}, 
\label{eq42}%
\end{equation}%
\begin{equation}
T_{\alpha\beta}^{\gamma}=
\dbra{\hat{F}_{\alpha}}
(\hat{\pi}_{\gamma}\otimes\hat{I})
\dket{\hat{F}_{\beta}}. 
\label{eq43}%
\end{equation}
We can further extend the above formulation to describe \textit{n}-qudit
operations. In this case, the bijective relation between $\mathcal{\hat{\hat
{S}}}$ and $\hat{\hat{\chi}}$ reads $\hat{\hat{\chi}}=\Lambda^{\otimes
n}(  \mathcal{\hat{\hat{S}}})  $ and $\mathcal{\hat{\hat{S}}%
=}\Lambda^{\otimes n}(  \hat{\hat{\chi}})  $. By using these
relations, we can straightforwardly extend Eqs. (\ref{eq40}) and (\ref{eq41})
to the bijection between the $\chi$- and $\mathcal{S}$-matrices for an 
\textit{n}-qudit operation.

It is obvious that not all the space of $\chi$- and $\mathcal{S}$-matrices
corresponds to physically realizable operations. For example, we can describe an 
anti-unitary operation by using the $\chi$- and $\mathcal{S}$-matrices, which
is evidently an unphysical operation. The requirement for the $\chi$- and
$\mathcal{S}$-matrices to represent physical quantum operations has been
extensively studied by many researchers. In the following, the requirements
common for the single- and two-qudit operations are summarized \cite{Arrighi}.

\begin{condition}
\label{Con1} 
(Hermiticity) The physical quantum operation $\mathfrak{S}$
should preserve Hermiticity; i.e., $\mathfrak{S}$ maps any Hermite operator
into an Hermite operator.
\end{condition}

\begin{condition}
\label{Con2} 
(Positivity) The physical quantum operation $\mathfrak{S}$ should
be positive; i.e., $\mathfrak{S}$ maps any positive operator into an positive operator.
\end{condition}

\begin{condition}
\label{Con3} 
(Complete positivity) The physical quantum operation
$\mathfrak{S}$ should be completely positive; i.e., positivity is preserved 
if we extend the L-space and HS-space by adding more qudits. That is, the
superoperator $\mathcal{\hat{\hat{S}}}\otimes\mathcal{\hat{\hat{I}}}$ on the
extended spaces should be positive.
\end{condition}
It is known that \ref{Con3} is sufficient for 
\ref{Con1} and \ref{Con2}, and \ref{Con2} is sufficient for 
\ref{Con1}. Therefore, we require complete positivity for a 
physical quantum operation. Complete positivity can be expressed as a
particularly simple condition for the $\chi$-matrix.

\begin{theorem}
\label{Th3} 
The linear operation $\mathfrak{S}$ is completely positive, iff
the $\chi$-matrix is positive.
\end{theorem}
This is natural on physical grounds, because the Choi operator $\hat{\hat
{\chi}}$ should be an unnormalized density operator 
associated with the system which was subjected to the quantum operation as 
will be discussed in the next section. 
In addition to \ref{Con3}, any physical
quantum operation should satisfy the following condition.

\begin{condition}
\label{Con4} 
(Trace non-increasing) The physical quantum operation
$\mathfrak{S}$ should be trace non-increasing; i.e., the mapped operator should
have trace less than one.
\end{condition}
This condition is simply expressed as the restriction on the $\chi$-matrix:
$\mathrm{Tr}_{1}\chi\leq I^{(1)}$ for a single-qudit operation and
$\mathrm{Tr}_{13}\chi\leq I^{(2)}$ for a two-qudit operation, where $I^{(n)}$ is
an identity matrix with size $d^{n}$.

It is needless to say that the $\chi$- and $\mathcal{S}$-matrices can be
defined with respect to arbitrary operator basis sets. Once these matrices are
given with respect to a particular operator basis set, they can be converted
into those defined with respect to the other basis set. It is obvious from Eqs.
(\ref{eq19}), (\ref{eq20}), (\ref{eq39}), and (\ref{eq38}) that the two
matrices defined with respect to different bases are unitarily equivalent. To be
specific, let $\chi^{E}$ and $\mathcal{S}^{E}$ be the $\chi$- and
$\mathcal{S}$-matrices for a single-qudit operation defined with respect to the
operator basis set $\{\hat{E}_{\alpha}\}_{\alpha=0}^{d^{2}-1}$ , and 
$\chi^{F}$ and $\mathcal{S}^{F}$ be those defined with respect to
the operator basis set $\{  {\hat{F}_{\alpha}}\}_{\alpha=0}^{d^{2}-1}$, 
where the two bases are unitarily related as shown in Eqs. (\ref{eq15}) and (\ref{eq16}). 
Then, these matrices should be written as
\begin{equation}
\mathcal{S}^{F}=\mathcal{U}^{\dag}\mathcal{S}^{E}\mathcal{U}, \label{eq44}%
\end{equation}%
\begin{equation}
\chi^{F}=\mathcal{U}^{\dag}\chi^{E}\mathcal{U}, \label{eq45}%
\end{equation}
where $\mathcal{U}\equiv\left[  \mathcal{U}{_{\alpha\beta}}\right]
_{\alpha=0}^{d^{2}-1}$ is a $d^{2}\times d^{2}$ unitary matrix. For the case
of two-qudit operation, we need to extend the set of the operator basis to cover
all the possible basis sets defined for the two-qudit operator space, 
that is, one that is a factorable set as well as not a factrorable set with
respect to the first and second systems. The general set of operator basis
$\{  {\hat{\Phi}_{\gamma}}\}_{\gamma=0}^{d^{4}-1}$ on
$\mathcal{H}_{d}^{\otimes2}$ should be unitarily related to the factorable 
operator bases ${\hat{E}_{\alpha}\otimes\hat{F}_{\beta}}$. If
we introduce a $d^{4}\times d^{4}$ unitary matrix $\mathcal{U}\equiv\left[
\mathcal{U}{_{\alpha\beta}}\right]  _{\alpha=0}^{d^{4}-1}$ that relates 
${\hat{\Phi}_{\gamma}}$ and ${\hat{E}_{\alpha}\otimes\hat{F}_{\beta}}$:
\begin{equation}
\hat{\Phi}_{\gamma}=\sum\limits_{\alpha,\beta=0}^{d^{2}-1}{\hat{E}_{\alpha
}\otimes\hat{F}_{\beta}}\mathcal{U}{_{\left[  {\alpha,\beta}\right]  \gamma},}
\label{eq46}%
\end{equation}
where $[\alpha,\beta]:=d^{2}\alpha+\beta$, it follows that Eqs. (\ref{eq44})
and (\ref{eq45}) also hold for two-qudit operations.

The $\chi$- and $\mathcal{S}$-matrices can be diagonalized by choosing the
appropriate operator basis sets, but are not necessarily diagonalized
simultaneously by a unique set. The operator basis set that
diagonalizes the $\chi$-matrix, each element of which is multiplied by the square root
of the associated eigenvalue, forms a particular set of Kraus operators in the
Kraus form of the quantum operation. Any set of Kraus operators can be
obtained by noting the unitary freedom in the Kraus form \cite{Nielsen}. It
follows from Eqs. (\ref{eq19}), (\ref{eq20}), (\ref{eq39}), and (\ref{eq38})
that the same operator basis set with the associated set of eigenvalues also
gives an operator-Schmidt decomposition for the \textit{L}-space superoperator
$\mathcal{\hat{\hat{S}}}$ \cite{Zyczkowski}. Therefore, the Kraus rank for the 
\textit{HS}-space superoperator $\mathcal{\hat{\hat{S}}}\left(  \odot\right)  $ 
and Schmidt number of the \textit{L}-space superoperator
$\mathcal{\hat{\hat{S}}}$ must be equal.

\section{Applications of $\chi$- and $\mathcal{S}$-matrices}

\label{applications}

This section presents the several applications of the $\chi$- and
$\mathcal{S}$-matrices. We discuss how these matrices and the present
formulation are useful for analysis and design of quantum operations.\\

\textbf{Experimental identification of quantum operations}

In the first example, we explain how useful the present formulation is for
experimental identification of quantum
operations \cite{Nambu,Altepeter,Mitchell,Martini,O'Brien,Secondi,Nambu2}. This
task is important because the development of any quantum device or circuit for
quantum computation and communication, which can be considered as an
input-output system that performs an intended quantum operation on its input
state and transforms it into its output state, necessarily requires experimental
benchmarking of its performance. The identification of a two-qudit device is
particularly interesting from a practical viewpoint as well as a scientific 
one because it may involve a nonseparable operation which has a purely quantum
mechanical nature, i.e., it cannot be simulated by using any classical method.

Identification of an input-output system amounts to identifying its $\chi$- or
$\mathcal{S}$-matrix, since these matrices characterize the system in question
completely as far as input and output data are concerned. The evaluated
matrices should reproduce the behavior of the system well enough when the
system is stimulated by any class of inputs of interest, and they 
should be useful for engineering the system of interest, e.g., to permit
control of the system, to allow transmission of information through the
system, to yield predictions of future behavior, etc. Identification problems
are commonly regarded as inversion problems, where the $\chi$- or
$\mathcal{S}$-matrix is to be statistically estimated from incomplete prior
knowledge of the system, using prior knowledge of corresponding inputs and the
collection of data obtained by measurement of outputs that usually contain
noise. In what follows, it will be shown that this common belief is not the
case for identification of quantum operations. To be specific, we can estimate
both the $\chi$- and $\mathcal{S}$-matrices without any inversion procedure
if we can make use of an entangled resource and a sequence of local
measurements assisted by classical communication \cite{Altepeter,Martini,Secondi,
D'Ariano2,Dur}.

Consider first the identification of a single-qudit operation. Equations
(\ref{eq19}) and (\ref{eq20}) show that all the elements of the $\chi$- and
$\mathcal{S}$-matrices are given by the expansion coefficients of the
operators $\hat{\hat{\chi}}$ or  $\mathcal{\hat{\hat{S}}}$ 
with respect to two different types of operator basis on
$\mathcal{H}_{d}^{\otimes2}$. Of these two operators, the Choi operator
$\hat{\hat{\chi}}$ is particularly useful since it is a positive operator
associated with the physical state of the bipartite object. To be specific, it
can be interpreted as the unnormalized output state from the system in
question where qudit 1 of the two qudits prepared in an isotropic state is
input into the system and undergoes the quantum operation $\mathfrak{S}$ while
qudit 2 is left untouched. Therefore, we can prepare the output state
corresponding to the normalized Choi operator ${{\hat{\hat{\chi}}/}d}%
\equiv\mathcal{\hat{\hat{S}}}\otimes\mathcal{\hat{\hat{I}}}(\hat{\rho}_{I})$
with the use of several copies of the isotropic-state input for the two
qudits. Thus, the identification of a single-qudit operation reduces to the
identification of a two-qudit state. It follows from Eq. (\ref{eq22}) that
every element of the $\mathcal{S}$-matrix can be directly obtained by
determining the expectation value of the corresponding product operator basis
$\langle {\hat{E}_{\alpha}\otimes\hat{E}_{\beta}^{\ast}}\rangle $
for the output states after the quantum operation has taken place. If the
basis ${\hat{E}_{\alpha}}$ is chosen to be the Hermitian operator basis
${\hat{\lambda}_{\alpha}}$, it suffices to make a set of $d^{4}$-independent
local measurements assisted by classical communication to determine the whole 
set of the real expectation values $\langle {\hat{\lambda}_{\alpha
}\otimes\hat{\lambda}_{\beta}}\rangle $. Accordingly, we can obtain the
$\mathcal{S}$-matrix defined with respect to the Hermitian operator basis set
$\{  {\hat{\lambda}_{\alpha}}\}  _{\alpha=0}^{d^{2}-1}$. Once the
$\mathcal{S}$-matrix is obtained, it is easy to convert it to the $\chi
$-matrix defined with respect to the same basis by using Eqs. (\ref{eq31}%
)-(\ref{eq33}) and also into the $\chi$- and $\mathcal{S}$-matrices defined with
respect to the arbitrary chosen basis by using Eqs. (\ref{eq44}) and
(\ref{eq45}).

The identification of a two-qudit operation can be carried out in the same way as the 
identification of a single-qudit operation. In this case, we prepare the state
corresponding to the Choi operator in Eq. (\ref{eq37}) with the use of several
copies of the product of isotropic states prepared in the four qudits. To
prepare the output state, we initially prepare the product of isotropic
states $\hat{\rho}_{I}^{(12)}\otimes\hat{\rho}_{I}^{(34)}$ in two pairs of two
qudits (qudits 1-2 and qudits 3-4). Then qudit 1 and qudit 3 are input into
the system in question, undergo the quantum operation $\mathfrak{S}$ jointly
while the other qudits are left untouched. This setup leads to an output
state ${{\hat{\hat{\chi}}/}d}^{2}$ in four-qudits. Thus, the identification of a 
two-qudit operation reduces to the identification of a four-qudit state. It
follows from Eq. (\ref{eq38}) that every element of the $\mathcal{S}$-matrix
for the two-qudit operation can be directly obtained by determining the real
expectation value of the corresponding product operator basis $\langle
{\hat{\lambda}_{\alpha}\otimes\hat{\lambda}_{\gamma}\otimes\hat{\lambda
}_{\beta}\otimes\hat{\lambda}_{\delta}}\rangle $ for the output states
after the quantum operation has taken place, if all the relevant basis sets
are chosen to be the Hermitian operator basis ${\hat{\lambda}_{\alpha}}$. It
suffices to make a set of $d^{8}$-independent local measurements assisted by
classical communication to determine the whole set of real expectation
values $\langle {\hat{\lambda}_{\alpha}\otimes\hat{\lambda}_{\gamma
}\otimes\hat{\lambda}_{\beta}\otimes\hat{\lambda}_{\delta}}\rangle $.
Accordingly, we can obtain the $\mathcal{S}$-matrix defined with respect to
the Hermitian operator basis $\{  {\hat{\lambda}_{\alpha}\otimes
\hat{\lambda}_{\beta}}\}  _{\alpha,\beta=0}^{d^{2}-1}$. The
$\mathcal{S}$-matrix can be converted into the $\chi$-matrix by using Eqs.
(\ref{eq40})-(\ref{eq43}). The $\mathcal{S}$- and $\chi$-matrices defined with
respect to arbitrary chosen bases can be obtained by applying the appropriate
matrix unitary transformation.\\

\textbf{Matrix analysis of quantum operations}

This section discusses in what way the $\chi$- and $\mathcal{S}$-matrices
contribute to developing quantum devices and circuits for quantum
computation and communication. We consider two classes of applications in
which these matrices offer useful mathematical models for quantum operations.
The first one concerns physical and information theoretic analysis of quantum
operations and the other concerns a logical calculus of quantum circuits or
algorithms comprised of a sequence of quantum operations.

Let us first consider the physical and information theoretic analysis of 
quantum operations. For this purpose, it is preferable to use the $\chi
$-matrix. This stems partly from the fact that the $\chi$-matrix is positive
and isomorphic to the density matrix in the doubled Hilbert space. Physically,
the diagonal elements of the process matrix show the populations of, and its
off-diagonal elements show the coherences between, the basis operators making
up the quantum operation, analogous to the interpretation of density matrix
elements as populations of, and coherences between, basis states. Owing to
Jamiolkowski isomorphism, the dynamic problems concerning quantum operations
can be turned into kinematic problems concerning quantum states in a 
higher dimensional space, and one can make use of a well-understood
state-based technique for analyzing the quantum operation. In what follows, we
show several illustrative examples and interesting problems from the physical
and information theoretic viewpoints.

The first example concerns the fidelity or distance measure between two
quantum operations. Several measures that make use of the above 
isomorphism have been proposed to quantify how close the quantum operation 
in question is to the ideal operation (usually a unitary operation) we are 
trying to implement. For example, the state fidelity defined between two states is 
extended to compare the two operations. The process fidelity $F_{p}$ is
defined by using the $\chi$-matrix $\tilde{\chi}$ of the system
in question and the rank one $\chi$-matrix $\chi_{ideal}$ of the ideal system
in the state-fidelity formula, that is, $F_{p}=\frac{1}{{d^{2n}}}{\text{Tr}%
}\tilde{\chi}\chi_{ideal}$, where \textit{n}=1 for single-qudit operation and
\textit{n}=2 for a two-qudit operation. The average gate fidelity $\bar{F}$
defined as the state fidelity between the output state after the quantum operation
and the ideal output can be calculated from the process fidelity. The purity
defined for the density matrix can be extended to characterize how much of a 
mixture the quantum operation introduces, which is also represented by the
simple function of the $\chi$-matrix of the system in
question \cite{O'Brien,White,Raginsky,Gilchrist}.

The next example concerns the analysis of a quantum operation acting on the
composite system. As mentioned before, the $\chi$-matrix of a two-qudit
operation is interesting from a practical as well as a scientific viewpoint.
The Jamiolkowski isomorphism for a two-qudit operation (Eqs. (\ref{eq37}) and
(\ref{eq38})) implies that the notion of entanglement can be extended from
quantum states to quantum operations. Analogously to what happens for states,
quantum operations on a composite system can be entangled \cite{Zanardi,Wang}. A
quantum operation acting on two subsystems is said to be separable if its
action can be expressed in the Kraus form
\begin{equation}
\mathcal{\hat{\hat{S}}}\left(  \odot\right)  =\sum\limits_{i}{(  {\hat
{A}_{i}\otimes\hat{B}_{i}})  \odot(  {\hat{A}_{i}\otimes\hat{B}%
_{i}})  ^{\dag},} \label{eq47}%
\end{equation}
where ${\hat{A}_{i}}$ and ${\hat{B}_{i}}$ are operators acting on each
subsystem \cite{Vedral,Barnum,Rains}. Otherwise, we say that it is nonseparable
(or entangled). Quantum operations that can be performed by local operations
and classical communications (the class of LOCC operations) are described by
separable quantum operations, yet 
there are separable quantum operations that cannot be implemented with LOCC
operations with probability one \cite{Bennett,Cirac,Harrow}. Anyway, these are
useless for creating entanglement in an initially unentangled system. It has been
pointed out by several authors that the separability and entangling properties
of quantum operations acting on two systems can be discussed in terms of the Choi
operator for two-qudit operations (Eq. (\ref{eq38})) \cite{Cirac,Dur,Harrow}. In
the present context, this reduces to discussing the separability properties of 
the $\chi$-matrices. For example, there is a condition for the $\chi$-matrix
equivalent to Eq. (\ref{eq47}): a quantum operation acting on two subsystems
is separable if its $\chi$-matrix can be written as $\chi=\sum\nolimits_{i}%
{\chi_{i}^{(A)}\otimes\chi_{i}^{(B)}}$, where ${\chi_{i}^{(A)}}$ and
${\chi_{i}^{(B)}}$ are the $\chi$-matrices for the quantum operation acting on
each subsystem. Thus, the separability of general quantum operations acting
on the composite system is reduced to the separability of its $\chi$-matrix. Since
the separability criterion and measure for the general $d^{4}\times d^{4}$
positive matrix is not fully understood, it remains as an important problem for
quantum information science to find such a criterion
and measure for general two-qudit quantum operations.

Let us turn our attention to a logical calculus of quantum circuits or
algorithms. For this purpose, the $\mathcal{S}$-matrix is practically useful.
This follows from the fact that \textit{L}-space superoperator algebra works
just like Dirac operator algebra. For example, consider the scenario in which
two-quantum operations $\mathfrak{S}_{1}$ and $\mathfrak{S}_{2}$ act
sequentially on a quantum system. Assume that the associated $\mathcal{S}%
$-matrices are given with respect to the same operator basis set $\{
\dket{\hat{E}_{\alpha}}\}
_{\alpha=0}^{N-1}$ in $\mathcal{L}_{N}$, where $N=d^{2}$ for single-qudit
operation and $N=d^{4}$ for two-qudit operation. Then the composite operation
$\mathfrak{S}=\mathfrak{S}_{1}\circ\mathfrak{S}_{2}$ is described by the
multiplication of \textit{L}-space superoperators
\begin{equation}
\mathcal{\hat{\hat{S}}}=\mathcal{\hat{\hat{S}}}_{1}\mathcal{\hat{\hat{S}}%
}_{2}
=\sum\limits_{\alpha,\beta=0}^{N-1}\mathcal{S}_{\alpha\beta}
\dketbra{\hat{E}_{\alpha}}{\hat{E}_{\beta}}, 
\label{eq48}%
\end{equation}
where 
\begin{equation}
\mathcal{S}_{\alpha\beta}=\sum\limits_{\gamma=0}^{N-1}{\left(  \mathcal{S}%
{_{1}}\right)  _{\alpha\gamma}\left(  \mathcal{S}{_{2}}\right)  _{\gamma\beta
}}. \label{eq49}%
\end{equation}
The extension to the case in which a sequence of a finite number of quantum
operations is applied to the same quantum system is trivial. Equation
(\ref{eq49}) implies that the $\mathcal{S}$-matrix of the composite operation
reduces to the multiplication of the $\mathcal{S}$-matrices of the individual
operations. This makes it practically advantageous using the $\mathcal{S}%
$-matrix to make a logical calculus of quantum circuits or algorithms comprised
of a sequence of elementary single- and two-qudit quantum operations.

Consider next the quantum circuit or algorithm comprised of a sequence of
quantum operations each of which acts not necessarily on the same quantum
system. This offers a general model for the quantum circuit acting on the large
numbers of qudits \cite{Aharanov}. To analyze
and design such a quantum circuit, we need to consider quantum operations
acting on the whole set of qudits and associated extended $\mathcal{S}%
$-matrix. Such an extended $\mathcal{S}$-matrix is non-trivial, but its
bijective $\chi$-matrix is trivially obtained by taking the tensor product with
identity matrix, that is, the $\chi$-matrix for the identity operation on the
irrelevant system. If the $\chi$-matrix for the quantum operation
$\mathfrak{S}$ on the relevant space is given by $\chi$, the $\chi$-matrix for
the extended quantum operation is given by $\chi\otimes I$. On the other hand,
we can trivially extend the conversion formulas (\ref{eq40}) and (\ref{eq41})
to those formulas for the quantum operation acting on more qudits. Therefore,
we can calculate the extended $\mathcal{S}$-matrix for each quantum operation
from the associated $\mathcal{S}$-matrix for the quantum operation acting on
the relevant space. The $\mathcal{S}$-matrix for a sequence of operations
acting on the space of the whole quantum systems can be calculated by
multiplication of the extended $\mathcal{S}$-matrices for the individual 
quantum operations.

The $\mathcal{S}$-matrix analysis of the quantum operation has the following
potential advantage. It can deal with a non-unitary operation in which mixed state
evolution occurs. Noisy quantum operation, probabilistic subroutines,
measurements, and even trace-decreasing quantum filters can be treated. This is 
in contrast to the usual analysis based on unitary matrix which can deal only 
with unitary gate in which only the pure state evolution is allowed. It thus 
offers us an mathematical model to analyze and design the logical operation of 
wider range of the complex quantum circuits and algorithms \cite{Aharanov}.

In the above discussion, we considered two applicational classes, i.e.,
physical and information theoretic analysis and logical calculus of
quantum operations, in which the $\chi$- and $\mathcal{S}$-matrices matrices
offer useful mathematical models. They have their own useful applications. The
present formulation will offer us the way of building bridges across the two 
applicational classes. For example, the entangling properties of the quantum
circuits comprised of a sequence of single- and two-qudit quantum operations
acting on several qudits can be discussed. The present formulation will also
help us to analyze and benchmark the quantum operation realized in the actual device.

\section{Conclusions}

\label{conclusions}

We have considered two matrix representations of single- and two-qudit quantum
operations defined with respect to an arbitrary operator basis, i.e., the $\chi$- and
$\mathcal{S}$-matrices. We have provided various change-of-representation
formulas for these matrices including bijections between the $\chi$- and
$\mathcal{S}$-matrices. These matrices are defined with the expansion coefficients of
two operators on a doubled Hilbert space, that is, the \textit{L}-space
superoperator and the Choi operator. These operators are mutually convertible through 
a particular bijection by which the Kronecker products of the relevant operator
basis and the dyadic products of the associated state basis are mutually converted.
From this fact, the mutual conversion formulas between two matrices are
established as computable matrix multiplication formulas. Extention 
to multi-qudit quantum operation is also trivial. These matrices are
useful for their own particular classes of applications, which might be
interesting from a practical as well as a scientific point of view.

We have
presented possible applications of the present formulation. By using the present
formulation, an experimental identification of a quantum operation can be
reduced to determining the expectation values of a Hermitian operator basis set
on a doubled Hilbert space. This can be done if we prepare several
copies of the isotropic-state input or the product of isotropic states input.
By using the $\chi$-matrix, we can make a physical as well as a quantum
information theoretic characterization of the quantum operation. In
particular, the $\chi$-matrix is useful to discuss the entangling properties
of the quantum operation acting on the composite system, since the problem of the
separability of the quantum operation is reduced to the problem of the
separability of the $\chi$-matrix. On the other hand, the $\mathcal{S}%
$--matrix is useful when we discuss the typical quantum circuit comprised of a
sequence of single- and two-qudit quantum operations each of which acts on
different quantum qubits. It is possible by considering the extended
$\mathcal{S}$--matrix of each quantum operation acting on the whole state
space of the relevant qudits. Such extended $\mathcal{S}$--matrices can be
calculated from the associated, bijective $\chi$-matrices by taking the tensor product
with the appropriate identity matrix. Accordingly, we can calculate the
$\mathcal{S}$--matrix for a quantum circuit by multiplying the
extended $\mathcal{S}$--matrices of each operation. This should be very useful to analyze and
design a wide range of the quantum circuits and algorithms involving
non-unitary operation.

We thank Satoshi Ishizaka and Akihisa Tomita for their helpful discussions. This 
work was supported by the CREST program of the Japan Science and Technology Agency.

\end{document}